%% file: main.tex
\documentclass[%reprint,
superscriptaddress,
 amsmath,amssymb,
 aps,
prl,
onecolumn,
]{revtex4-2}

\usepackage{xpatch} 
\makeatletter
\xpatchcmd\@collaboration@present{(}{\medskip}{}{}
\xpatchcmd\@collaboration@present{)}{}{}{}
\renewcommand{\@fnsymbol}[1]{%
  \ifcase#1\or *\or \dagger\or \ddagger\or
  \mathsection\or \mathparagraph\or \|\or **\or \dagger\dagger \or \ddagger\ddagger
  \else\@ctrerr\fi}
\makeatother

\newcommand{\thor}{$^{229}$Th}

\newcommand{\lisaf}{LiSrAlF$_6$}

\newcommand{\thox}{$^{229}$ThO$_2$}
\newcommand{\thoria}{ThO$_2$}

\usepackage[dvipsnames]{xcolor}
\definecolor{ricky}{cmyk}{0, 0.7808, 0.4429, 0.1412}

\usepackage{siunitx}
\usepackage{braket}
\usepackage{graphicx}
\usepackage[colorlinks=true, allcolors=blue]{hyperref}
\usepackage{gensymb}
\usepackage{upgreek}
\usepackage{tikz}

\usepackage{float}  
\usepackage{xcolor}

\definecolor{darkgreen}{rgb}{0.0, 0.5, 0.0}  % dark green definition
\newcommand{\apd}[1]{\textcolor{darkgreen}{#1}}

\usepackage{braket}
\usepackage{graphicx}

\def\thf{\thor F$_4$ }

\usepackage[colorlinks=true, allcolors=blue]{hyperref}
\usepackage[version=3]{mhchem}
\usepackage{physics,multirow}
\usepackage{tikz}
\usetikzlibrary{calc,decorations.pathmorphing,decorations.markings,arrows.meta}
\tikzset{>={Stealth[scale=2]}} % enlarge arrowheads
\tikzset{mid arrow/.style={decoration={markings,mark=at position 0.5 with {\arrow{>}}},postaction={decorate}}}
\begin{document}

\title{\thor{} Nuclear Spectroscopy in an Opaque Material: Laser-Based Conversion Electron M\"{o}ssbauer Spectroscopy of \thox}

\author{Ricky Elwell\footnote{These authors contributed equally to the work.}}
\affiliation{Department of Physics and Astronomy, University of California, Los Angeles, CA 90095, USA}
\author{James E. S. Terhune\footnotemark[1]}
\affiliation{Department of Physics and Astronomy, University of California, Los Angeles, CA 90095, USA}
\author{Christian Schneider\footnotemark[1]}
\affiliation{Department of Physics and Astronomy, University of California, Los Angeles, CA 90095, USA}
\author{Harry W. T. Morgan}
\affiliation{Department of Chemistry, University of Manchester, Oxford Road, Manchester M13 9PL, UK}
\author{Hoang Bao Tran Tan}
\affiliation{Department of Physics, University of Nevada, Reno, Nevada 89557, USA}
\affiliation{Computational Physics Division, Los Alamos National Laboratory, P.O. Box 1663, Los Alamos, New Mexico 87545, USA} 
\author{Udeshika C. Perera}
\affiliation{Department of Physics, University of Nevada, Reno, Nevada 89557, USA}
\author{Daniel A. Rehn}
\affiliation{Computational Physics Division, Los Alamos National Laboratory, P.O. Box 1663, Los Alamos, New Mexico 87545, USA}
\author{Marisa C. Alfonso}
\affiliation{Eckert \& Ziegler Analytics, Inc., Atlanta, Georgia 30318, USA}
\author{Lars von der Wense}
\affiliation{Department of Physics, Johannes Gutenberg-Universit\"{a}t Mainz, Mainz 55128, Germany}
\author{Benedict Seiferle}
\affiliation{Faculty of Physics, Ludwig-Maximilians-Universit\"{a}t M\"{u}nchen, 85748  Garching, Germany}
\author{Kevin Scharl}
\affiliation{Faculty of Physics, Ludwig-Maximilians-Universit\"{a}t M\"{u}nchen, 85748  Garching, Germany}
\author{Peter G. Thirolf}
\affiliation{Faculty of Physics, Ludwig-Maximilians-Universit\"{a}t M\"{u}nchen, 85748 Garching, Germany}
\author{Andrei Derevianko}
\affiliation{Department of Physics, University of Nevada, Reno, Nevada 89557, USA}
\author{Eric R. Hudson}
\affiliation{Department of Physics and Astronomy, University of California, Los Angeles, CA 90095, USA}
\affiliation{Challenge Institute for Quantum Computation, University of California Los Angeles, Los Angeles, CA, USA}
\affiliation{Center for Quantum Science and Engineering, University of California Los Angeles, Los Angeles, CA, USA}

%\date{\today} 
\date{\today} % Leave empty to omit a date

%\begin{abstract}

%\end{abstract}

\maketitle

\textbf{The exceptionally low-energy \ce{^229Th} nuclear isomeric state is expected to provide a number of novel and powerful applications~\cite{Tkalya1996, Peik_2003}, including the construction of a robust and portable solid-state nuclear clock~\cite{Rellergert2010a}, perhaps contributing to a redefinition of the second~\cite{Morgan2025ThSO42}, exploration of nuclear superradiance~\cite{Dicke1954,Liao2012},  and tests of fundamental physics~\cite{Flambaum2006a,Flambaum2009,Fuchs2025,Atypas2022}. 
Further, analogous to the capabilities of traditional M\"{o}ssbauer spectroscopy,  the sensitivity of the nucleus to its environment can be leveraged to realize \emph{laser} M\"{o}ssbauer spectroscopy and with it new types of strain and temperature sensors~\cite{Rellergert2010a, Higgins2025} and a new probe of the solid-state environment~\cite{morgan_internal_conversion_2024}, all with exquisite sensitivity. 
However, current paradigms for probing the nuclear transition in a solid require the use of a high band gap, vacuum-ultraviolet transmissive host, severely limiting the applicability of these techniques – e.g. so far laser excitation has only been demonstrated in the high band gap metal fluoride systems \ce{CaF2}~\cite{Tiedau2024}, \ce{LiSrAlF6}~\cite{Elwell2024}, and \ce{ThF4}~\cite{zhang2024thf}. 
Here, we report the first demonstration of laser-induced conversion electron M\"{o}ssbauer spectroscopy, which provides the ability to probe the nuclear transition in a material that is opaque to light resonant with the nuclear transition.
Specifically, we excite the nuclear transition in a thin \ce{ThO_2} sample whose band gap ($\sim6$~eV) is considerably smaller than  the nuclear isomeric state energy (8.4~eV).
As a result, the excited nucleus can quickly decay by internal conversion~\cite{Tkalya2000}, resulting in the ejection of electrons from the surface. 
By collecting these conversion electrons, nuclear spectroscopy can be recorded.
Unlike fluorescence spectroscopy, this technique is compatible with materials whose work function is less than the nuclear transition energy, opening a wider class of systems to study. 
Further, because \thoria{} can be made from spinless isotopes~\cite{Morgan2025ThSO42} and the internal conversion decay process reduces the isomeric state lifetime to only $\sim 10~\mu$s, allowing $\sim 10^8$ relative reduction in clock interrogation time, a conversion-electron-based nuclear clock could lead to a $\sim 10^4$ improvement in clock instability. 
}

The recent observation~\cite{Tiedau2024, Elwell2024} of direct laser excitation of the \thor{} isomeric state in high band gap crystals has, after almost 50 years of work, opened the door to a laser-accessible nuclear transition and driven rapid progress in the development of solid-state optical clocks. 
Already, the nuclear transition frequency has been compared to an atomic clock and its linewidth studied~\cite{Zhang2024-Th229Comb}, and work has begun to optimize the clock performance by developing both a theoretical understanding of nuclear quenching mechanisms~\cite{morgan_internal_conversion_2024} and using them to shorten the clock interrogation cycle~\cite{Terhune2024-Photoquenching,Schaden2024-Th-quenching}.
Similarly, excitation of the nuclear transition in a vacuum ultraviolet (VUV) transmissive thin film has opened the door to integrated-photonic-based nuclear clocks and sensors~\cite{zhang2024thf}. 
Further, new high bandgap materials have been analyzed that could provide dramatic simplification of the clock by, e.g., doping \thor{} into a nonlinear optical crystal~\cite{Morgan2024-Th229NonLinear} and, by using principles of molecular design, provide a system with the potential for performance orders of magnitude beyond any current or planned optical clock~\cite{Morgan2025ThSO42,Morgan2025polyatomic}. 

Because the \thor{} nucleus provides a highly-controllable system which can be deployed into various hosts, it is expected, in analogy to M\"{o}ssbauer spectroscopy, that these same techniques can also be used as new probes of the solid-state chemical and nuclear environment. 
However, in all studies to date, nuclear excitation is detected by observing the resulting nuclear fluorescence, which requires the material band gap to be larger than the isomeric transition energy ($>E_\text{iso}$).
This requirement severely limits the material hosts available for study and, therefore, it is highly desirable to extend the spectroscopy of the isomeric transition to low band gap ($<E_\text{iso}$) environments.

Conversion electron M\"{o}ssbauer spectroscopy (CEMS) of \thor{} has been proposed as a method to extend the technique of laser-based nuclear spectroscopy to low band gap materials~\cite{vonderwense2017, vonderWense2019_cems_proposal}. 
CEMS utilizes the fact that if the nuclear energy is larger than the material band gap, the internal conversion (IC) relaxation process is possible by promoting an electron across the band gap~\cite{Tkalya2000,Tkalya2015}. 
If this promoted IC electron originates in a shallow-enough state in the valence band, the electron can overcome the work function barrier and emerge from the material surface. 
By detecting these conversion electrons, nuclear spectroscopy can be recorded. 

In addition to providing a means to study low band gap materials, CEMS could also allow significant improvements in nuclear clock stability as the IC decay rate is roughly $10^8$ times faster than the radiative decay rate, enabling a much faster clock interrogation rate, leading to a projected clock instability of $\sim 10^{-18}$ at 1~s. 
Further, a CEMS-based nuclear clock could operate by simply reading out the CEMS photocurrent, providing a means to greatly simplify and miniaturize future nuclear clocks~\cite{Hudson2025CEMSpatent}. 

Here, we report the first demonstration of laser-based CEMS of any nucleus.
Specifically, a VUV laser system is used to excite \thor{} nuclei in a thin sample of thoria (\thoria{}).
\thox{} was chosen as the first host as it has a low band gap ($\sim$6 eV~\cite{tho2_band_gap}) and is readily available as a stochiometric Th compound. 
Conversion electrons from the \thox{} are detected as a function of laser excitation energy providing the first laser-based CEMS nuclear spectrum, which determines the \thor{} nuclear transition frequency as $2020407.5(2)_\textrm{stat}(30)_\textrm{sys}$~GHz, consistent with previously reported values of the transition~\cite{Tiedau2024,Elwell2024,zhang2024thf, Zhang2024-Th229Comb}, and whose width is consistent with the laser linewidth of our system reported in previous studies~\cite{Elwell2024,zhang2024thf}.
Further, the IC lifetime is measured as $\sim 10~\mu$s, which is consistent with that measured in \thor{} implantation studies~\cite{seiferle2017_lifetime_IC} and theoretical calculations of the IC lifetime described below.

In what follows we describe the \thox{} target, experimental apparatus, and the spectroscopic protocol, before presenting the results of the CEMS spectroscopy. 
%\EricComment{Theory?}
Next, we present calculations of the isomer shift and IC lifetime in \thox{} using a theoretical approach that combines solid-state density functional theory and relativistic atomic many-body perturbation theory methods.
Following these results, we use the measured lifetime of the internal conversion decay of the isomeric state in \thox{} to evaluate the ideal performance of a \thox{} conversion electron clock, and find its projected instability to be $\sim$2$\times 10^{-18}$ at 1~s.

\subsection{\thox{} Target and Apparatus}
\begin{figure*}[t!]
    \centering
    \includegraphics[width=0.99\textwidth]{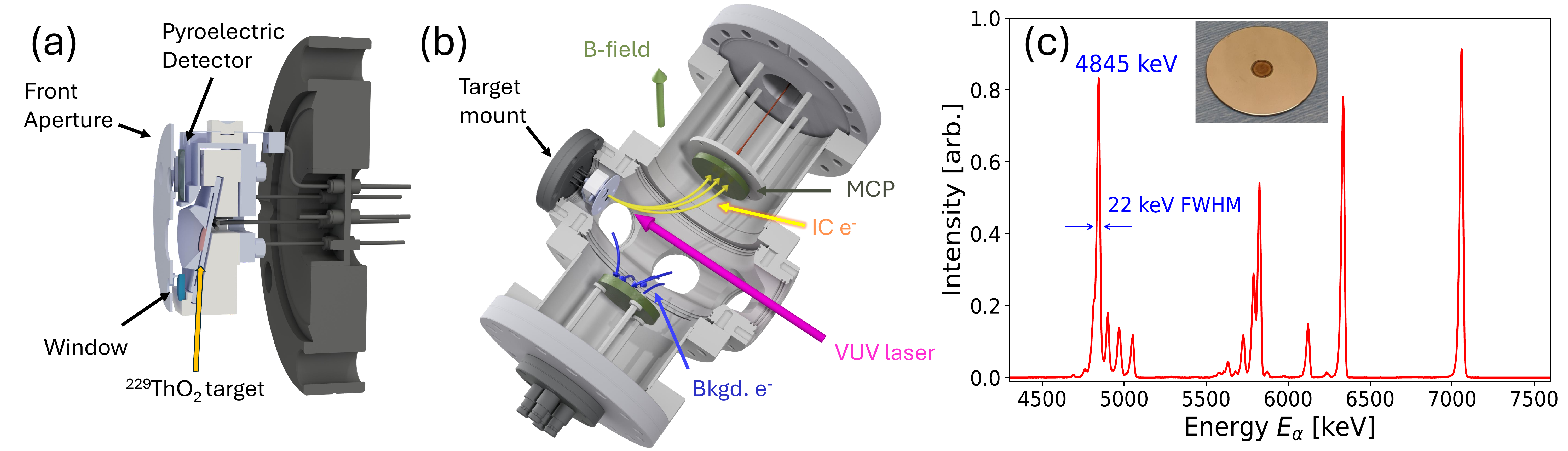}
    \caption{(a)~Cut-away rendering of the \thox{} target mount. Arrows denote front aperture, window, target, and pyroelectric detector. (b)~Rendering of the spectroscopy chamber. (Magenta arrow) Direction of VUV laser propagation. (Yellow arrows) IC electron trajectories from target to detection MCP. (Blue arrows) Background photoelectrons generated from VUV scatter diverted to secondary electrode. (Green arrow) Direction of static B-field used to guide IC electrons. (c)~$\alpha$-spectrum of the \thox{} target. (Inset) Photograph of \thox{} target used in this study. The peak labeled to 4845 keV corresponds to the dominant $\alpha$-decay mode of \thor{}. The other large peaks correspond to the $\alpha$-decays of daughter nuclei. The \thor{} peak had a FWHM of $\sim$22 keV, consistent with energy loss through a $\sim$10 nm sample, as estimated with SRIM~\cite{Ziegler2010-srim}.}
    \label{fig:thoria_target_mount}
\end{figure*}

The CEMS target was constructed by electrodepositing \thox{} onto a stainless steel disk (see Fig.~\ref{fig:thoria_target_mount} for image of target, procured from Eckert \& Ziegler Analytics, Inc.). Although other oxide and hydroxide compounds of Th may be present in the target, this paper assumes that the predominant chemical form of \thor{} in the target is \thox{}, based on the method of target preparation (see SI for details).
The diameter of the \thox{} is $\approx 5$~mm and its thickness is estimated from the width of the $\alpha$-particle spectrum as $\sim 10$~nm (Fig.~\ref{fig:thoria_target_mount}(c)), which is consistent with a measured total activity of $\approx$6.3 kBq. 
The VUV absorption of \thoria{} is estimated as $\alpha \approx 0.1$~nm$^{-1}$~\cite{tho2_abs_ellipsometry}, implying that \thor{} nuclei within a depth of $\approx 10$~nm can be efficiently excited by the laser. 
In addition, the inelastic mean free path of electrons in solids at the eV energy scale is typically $\sim 10$~nm~\cite{seah_dench_imfp_electrons}, meaning that an IC electron generated at greater depths likely scatters before extraction from the surface. 
As such, further \thox{} thickness primarily contributes to the background in the form of electrons produced via radioactive decays~\cite{stellmer2018_exoelectrons}, and thus a thickness of $\approx 10$~nm is likely optimal for CEMS.

CEMS spectroscopy is recorded by directing a tunable VUV laser onto the \thox{} target.
Briefly, VUV radiation was produced via resonance-enhanced four-wave mixing of two pulsed dye lasers in Xe gas. 
The frequency of the first pulsed dye laser, $\omega_u$, was locked to the $5p^{6 ~1}S_0~\rightarrow~5p^5\left(^2P^{\circ}_{3/2} \right) 6p~^2\left[1/2\right]_0$ two-photon transition of Xe at $\sim$ 249.63~nm. 
The frequency of the second pulsed dye laser, $\omega_v$, was scanned to produce VUV radiation in the Xe cell given by the difference mixing relation $\omega = 2\omega_{u} - \omega_v$. 
All three laser beams then impinge off-axis with respect to a MgF$_2$ lens, whose chromatic dispersion is used with downstream pinholes to spatially filter the VUV beam and pass it towards the \thox{} spectroscopy chamber. 
The laser system delivers 30 pulses per second to the target with a typical VUV pulse energy of $\sim 6-8$ \SI{}{\micro J}/pulse (see Refs.~\cite{JeetThesis2018, Elwell2024} for details).
For these experiments, \ce{N2} gas is added to the Xe cell to quench a process leading to amplified spontaneous emission at 147~nm (see SI for details), which occurs on a timescale similar to IC decay and otherwise contributes to the photoelectron background. 

The \thox{} target is held in a custom mount (see Fig.~\ref{fig:thoria_target_mount}(a)) that provides the ability to electrically bias the sample, as well as VUV laser monitoring capabilities. 
Biasing of the target is necessary as the VUV laser causes a prompt burst of photoelectrons when it impinges on the \thox{} target, which will overwhelm the electron detector. 
These initial photoelectrons are suppressed for the first $\sim$100 ns after the laser pulse by positively biasing the \thox{} target to 135~V, before the bias voltage is changed to -405~V to aid the extraction of IC electrons from the surface; at all times the front aperture of the mount is held at 100~V.
The mount also houses a fused silica target that fluoresces under VUV illumination aiding alignment and a bare pyroelectric crystal to provide \emph{in situ} monitoring of the VUV laser pulse energy. 
The target chamber is mounted on a combined lever arm-stepper motor system that allows it to be raised and lowered, causing the VUV laser to illuminate either the \thox{} target or laser energy monitor. 
To limit hydrocarbon buildup on the target and to limit scattering of the IC electrons due to background gases, the target chamber was ozone plasma cleaned prior to baking, and operated at a pressure of $\sim$10$^{-7}$ Pa. 

In principle, IC electron detection can be accomplished by simply positively biasing, e.g., a multi-channel plate (MCP) detector held near the \thox{}.
However,  it was found that scattered light from the VUV laser generated a large afterglow of photoelectrons originating at positions dispersed throughout the chamber which lasted tens of microseconds – presumably due to fluorescence of components within the beamline and chamber.
To overcome this background, a combination of electric and magnetic fields were used that focused electrons originating from the \thox{} target onto a detection MCP, while diverting electrons generated elsewhere to another region of the chamber (see SI).
Finally, as an additional layer of protection for the detection MCP, its front plate voltage was controlled so that it had no gain during the initial burst of photoelectrons from the target.

\subsection{Spectroscopy}
\begin{figure}
    \centering
    \begin{tikzpicture}
        \node[anchor=south west, inner sep=0] (image) at (0,0) {
            \includegraphics[width=0.49\linewidth]{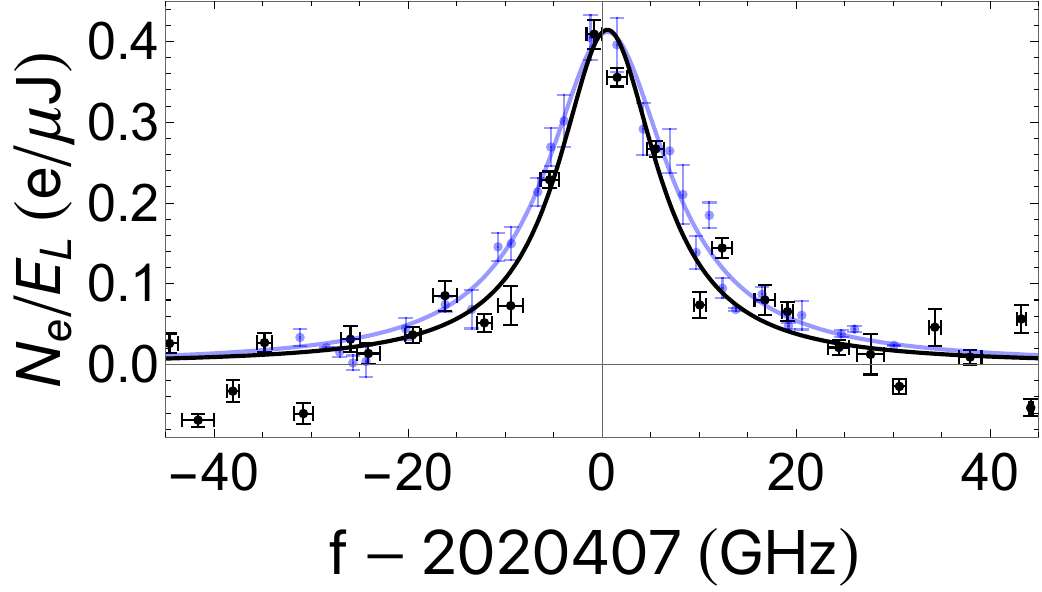}
            \includegraphics[width=0.495\linewidth]{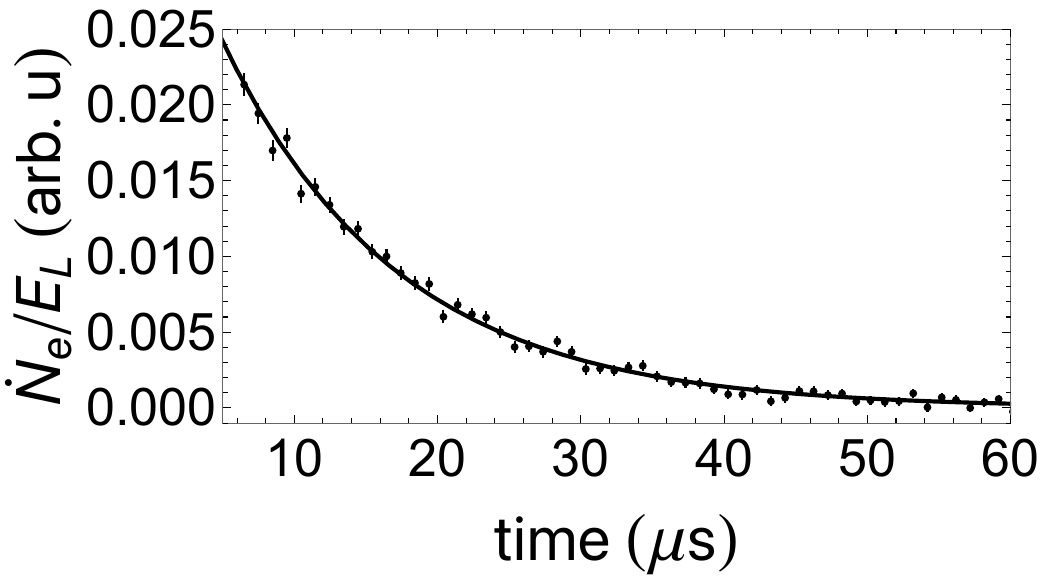}
        };
        \node at (1.8,4.6) {\textbf{{\fontfamily{ptm}\selectfont
\large (a)}}};  
        \node at (17.2,4.6) {\textbf{{\fontfamily{ptm}\selectfont
\large (b)}}};
    \end{tikzpicture}
    \caption{(a) The laser-based CEMS spectra of \thox{} is shown in black with a fit of a Lorentzian profile. Each point is an average of $\geq 140,000$ laser pulses. A constant baseline of $\approx$ 75 e$^-$ per laser pulse has been subtracted. The vertical error bars denote the standard error, while the horizontal error bars are the standard deviation of the laser frequency for the points in the bin. For comparison, the radiative decay spectrum of \ce{^229Th}:\ce{LiSrALF6}, taken from Ref.~\cite{Elwell2024}, is shown in blue.  (b) The IC decay rate as a function of time is shown in black alongside a fit of a decaying exponential with a background, leading to an IC lifetime of $12.3(3)~\mu$s.}
    \label{fig:CEMS_data}
\end{figure}

Using this apparatus, CEMS was performed by recording the number of detected electrons in the window between $6~\mu$s and $40~\mu$s after each laser pulse as a function of VUV laser wavelength; this window was chosen based on previous theoretical estimates~\cite{Tkalya2015} and observations of IC decay~\cite{VonderWense2016a} following \ce{^233U} decay.
The electron signal, normalized by the laser pulse energy, is shown in Fig.~\ref{fig:CEMS_data}(a) as a function of frequency for a region $\sim100$~GHz wide and centered on the \thor{} nuclear transition energy. 
These data were recorded by firstly measuring the pulse energy, then detecting the total number of electrons emitted after 36,000 pulses (20 minutes) before measuring the pulse energy again.
This process was repeated to record a laser CEMS spectrum.
The data in Fig.~\ref{fig:CEMS_data}(a) is the average of four spectra with the vertical error bars representing the standard error.
Due to differences in system alignment between campaigns, the photoelectron background varied slightly between different spectra. 
Therefore, a Lorentzian with a background fit was used to determine and subtract the photoelectron background before the data is combined. 
Further, because the laser frequency is not perfectly controlled these spectra are binned in frequency and the horizontal error bars are the standard deviation of the frequency. 
Each data point represents the average of at least 140,000 laser pulses. 

A nonlinear, least squares fit of a Lorentzian to the spectrum yields a central value of $2020407.5(2)_\textrm{stat}(30)_\textrm{sys}$~GHz and full-width-half-maximum linewidth of 12.4(4)~GHz, both consistent with our observations in \ce{LiSrAlF6}~\cite{Elwell2024} and \ce{ThF4}~\cite{zhang2024thf}. 
Here, statistical error is the 68\% confidence interval and the systematic uncertainty is predominantly due to the accuracy of the wavemeter and likely conservative. 
On resonance, an average of 0.41(5)~$e$/$\mu$J per laser pulse is detected.
The expected signal can be estimated as $\eta_\text{e}\eta_\text{c} N_e \times e^{-t_i/\tau_\text{IC}}$, where $\eta_\text{e}$ is the extraction of efficiency of IC electrons from the sample, $\eta_\text{c}$ is the collection efficiency of electrons emitted from the sample, $N_e$ is the number of excited \thor{} nuclei, and $t_i$ is the time that electron counting begins after the laser pulse. 
Based on calibrations of the apparatus (see SI), the expected signal is found as $0.15_{-0.09}^{+0.25}$ $e^-$/$\mu$J, in reasonable agreement with the recorded spectrum.

Using the same system, the lifetime of the IC decay is measured by comparing the time binned electron counts collected on resonance in the first $190~\mu$s after laser excitation. 
Specifically, the average number of time binned electron counts per $\mu$J of laser energy is obtained for both the on and off ($\sim100$~GHz detuning) resonant illumination periods. 
The average off resonance counts are then subtracted from the average on resonance counts to remove any photoelectron background, and the result is plotted in Fig.~\ref{fig:CEMS_data}(b). 
A background free nonlinear least-squares fit of a decaying exponential reveals an IC lifetime of 12.3(3)~$\mu$s, as shown in Fig.~\ref{fig:CEMS_data}(b). 
As discussed later, the IC decay rate is sensitive to the local chemical environment, so a range of decay rates is possible and this lifetime should be interpreted as an estimate. 

\subsection{Calculation of Isomer Shifts and Internal Conversion Lifetime in \thox{}}
The measured isomer energy is in agreement with the earlier spectroscopic experiments in three wide band gap hosts~\cite{Tiedau2024,Elwell2024,zhang2024thf}. 
Theoretically, this is supported by the calculations~\cite{perera2025-isomer-shift-Th229} of isomer shifts for a large variety of \thor{} solid-state hosts and by our \thox{}-specific calculations presented in the SI. 
For example, we expect that the isomer shift between bulk \thox{} and \thor{}:\lisaf{} is on the order of 100 MHz,
well below the $\sim 3\, \text{GHz}$ reported experimental resolution.
Our calculations predict the nuclear transition frequency for bulk \thox{} to be $2,\!020,\!407,\!338(70) \, \text{MHz}$.

\begin{figure}[t!]
    \centering
    \includegraphics[width=1\linewidth]{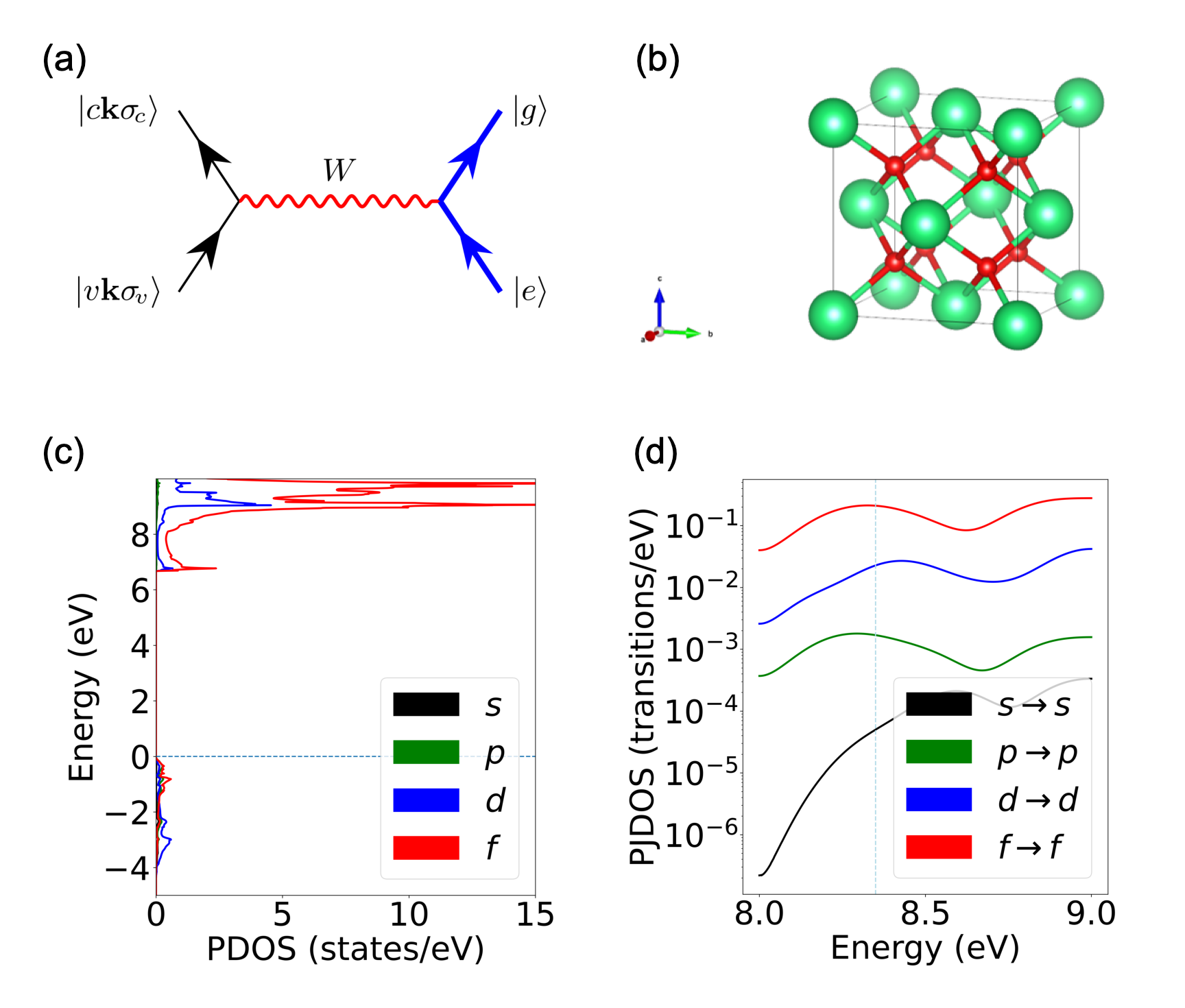}
    \caption{(a)  Feynman diagram of the internal conversion process. The interaction is mediated by a virtual photon exchange, corresponding to the hyperfine interaction $W$. The crystal momentum $\mathbf{k}$ is conserved in this process.  (b) Conventional unit cell of \ce{ThO2} with Th atoms in green and O atoms in red. (c) Th projected density of states (PDOS) computed with $G_0W_0$. The horizontal dashed line denote the Fermi level. (d) Th projected joint density of states (PJDOS) computed with $G_0W_0$. The vertical dashed line denotes the nuclear isomeric energy.}
    \label{fig:ThO2 structure and PDOS}
\end{figure}

In the IC process, the energy of the excited nucleus is transferred to the electrons.
In a crystalline solid, that means promoting a valence band electron $\ket{v \mathbf{k}\sigma_v}$ into a conduction band state $\ket{c \mathbf{k} \sigma_c }$ and creating a hole in the valence band (Fig.~\ref{fig:ThO2 structure and PDOS}(a)).
The process is mediated by the hyperfine interaction (HFI) $W$ which, for periodic lattices, conserves the electronic crystal momentum $\mathbf{k}$, but can connect electronic states of different spin projections $\sigma_{v,c}$.
We derived the rate for the described IC process (see SI) as 

\begin{align}
 \Gamma_\text{IC} =  
  \frac{2 \pi}{\hbar }  \,  
  \frac{1}{2I_e+1} \sum_{M_gM_e} \sum_{\sigma_c\sigma_v} 
\overline{|{W^{ge}_{c\sigma_c v \sigma_v}(\mathbf{k})}  |^2} G_{cv}(\hbar\omega_\text{nuc}) \,.
\label{Eq:ICRate}
\end{align}
Here $\ket{e}$ and $\ket{g}$ are the isomer and the ground nuclear states with spins $I_{e,g}$ and magnetic quantum numbers $M_{e,g}$.
The HFI $W$ connects the two nuclear states and the valence and conduction band electronic states. 
$G_{cv}(\hbar\omega_\text{nuc})$ is the conventional~\cite{callaway1991Book} Joint Density of States (JDOS) at the nuclear transition energy $\hbar\omega_\text{nuc}$. 
The bar denotes averaging over a surface of constant energy in $\mathbf{k}$-space. 

The same JDOS appears in the theory of electromagnetic interband absorption~\cite{callaway1991Book} and is clearly a highly material-dependent quantity. Therefore, to understand the physics of the IC process, and in particular its rate, we calculated the properties of \thox{} with periodic electronic structure theory ($GW$ and DFT).
The calculations used the $G_0W_0$ approximation for the electronic self‐energy and the Bethe–Salpeter Equation (BSE) for optical properties (see SI for details). 
The fundamental band gap computed with $G_0W_0$ is 6.2 eV, in good agreement with experimental values~\cite{Mock2019,Dugan2022} 
(calculation details provided in the SI).

The absorption spectrum, computed with $G_0W_0$+BSE, agrees with the measurements in Ref.~\cite{tho2_abs_ellipsometry}, supporting our estimation of $\alpha \approx 0.1$~nm$^{-1}$.

\ce{ThO2} crystallizes in the fluorite structure (space group $Fm\bar 3m$, \#225), where each Th atom is coordinated by eight oxygen atoms in a cubic arrangement (Fig.~\ref{fig:ThO2 structure and PDOS}(b)). 
The Th$^{4+}$–O$^{2-}$ bonding is predominantly ionic with a modest covalent admixture arising from overlap of Th $6d/5f$ and O $2p$ orbitals. That Th is in the +4 oxidation state may be verified from the projected density of states (PDOS) on Th, where the $5f$ and $6d$ components in the valence band are nearly empty. In contrast, the Th $5f$ and $6d$ valence orbitals form large peaks in the conduction band, indicating the transfer of an electron from an \ce{O^{2-}} anion to form Th${}^{3+}$. Both the valence and conduction bands contain small contributions from Th-projected $p$ orbitals. As it turns out, the IC process primarily involves transferring electrons between these $p$-orbital components.

In computing the HFI matrix element in Eq.~\eqref{Eq:ICRate}, it is convenient to project the valence and conduction band states $\ket{v \mathbf{k}\sigma_v}$ and $\ket{c \mathbf{k} \sigma_c }$ onto the basis of Th atomic orbitals. 
The matrix element ${W^{ge}_{c\sigma_c v \sigma_v}(\mathbf{k})}$ may then be expressed as a sum over different HFI matrix elements connecting atomic orbitals of definite angular momenta, weighted by the expansion coefficients of $\ket{v \mathbf{k}\sigma_v}$ and $\ket{c \mathbf{k} \sigma_c }$.
To the lowest order, cross terms arising when squaring the expansion of ${W^{ge}_{c\sigma_c v \sigma_v}(\mathbf{k})}$ can be neglected. 
This allows the expression of the IC rate in terms of the nuclear ground state HFI constants $A$ and the ``projected joint density of states'' (PJDOS) in Th.
The PJDOS is a JDOS weighted by the projections of the crystal electronic states onto the thorium atomic orbitals that allows us to describe the character of the delocalized crystal orbitals close to the thorium nucleus (see SI).
The HFI $A$ constants for Th$^{3+}$ are known from experiments and high-precision relativistic atomic-structure calculations, see Ref.~\cite{morgan_internal_conversion_2024}. 
The PJDOS of transitions with the largest contributions to the IC rate are shown in Fig.~\ref{fig:ThO2 structure and PDOS}(c). 

Using these values, we arrive at an estimate for the IC rate of 
 $   \Gamma_{\rm IC} \approx 1.3\times10^4\,{\rm s}^{-1}$, 
corresponding to an IC lifetime of $\approx 80~\mu$s.

Sources of the disparity between the measured and theoretical lifetimes could include known errors in the local projections of the plane wave orbitals done by VASP~\cite{Maintz2013}, 
small changes to the valence and conduction states expected with the addition of spin-orbit coupling in solid-state calculations, cross- and higher-order expansion terms neglected in our evaluation of the rate~\eqref{Eq:ICRate}, and deviations of the sample from bulk \thox{} due to both surface and self-radiation damage effects. Compared to the pristine \thox{}, the IC rate for \thor{} adjacent to point defects can be enhanced  by the electric quadrupole HFI contribution~\cite{Bilous2018} due to symmetry breaking.
Nevertheless, the order of magnitude agreement supports the physical interpretation that observed IC decay results from the nucleus relaxing via transferring an electron from an oxide anion into the Th $6p$ component of the conduction band.

\subsection{Projected Clock Performance}
In addition to its use as a new chemical probe, laser-based CEMS may allow the construction of a new type of nuclear clock with several advantages over crystal-based clocks~\cite{vonderWense2020_concept_cems}.
Chief among these advantages are a greatly reduced clock interrogation time, as the IC decay rate is about $10^8$ times faster than radiative decay, and the potential for clock readout by simply monitoring the current leaving the target, which could facilitate dramatic miniaturization. 

For such a clock, assuming \thox{} is produced from \ce{^16O}, the largest sources of instability would be broadening due to magnetic dipole interactions between the \thor{} nuclei and lifetime broadening. 
Given that the Th-Th distance in \thoria{} is 3.96 $\AA{}$~\cite{belle_berman_1984_tho2_report}, the expected Zeeman broadening due to to neighboring \thor{} nuclei is $\sim 10$~Hz.
Thus, the primary source of instability is the lifetime broadening of $\Gamma \approx $ 2$\pi \times$ 16 kHz. 
Assuming a polished face of a $\sim 10$~nm thick single crystal is prepared and realizes an electron extraction efficiency $\eta_\text{e} \approx 0.5$~\cite{seah_dench_imfp_electrons}, and that the photoelectron background is eliminated by material purification and baffling, a 100 $\mu$W laser leads to a projected clock instability of $\sim2\times 10^{-18}$ at 1~s averaging. 
Interestingly, it is possible that the electric current from the IC process could also be used as a form of readout, as it would generate a current of $\sim$300 nA on the $\mu$s timescale, providing a simple means for clock operation and locking~\cite{Hudson2025CEMSpatent}. 

\section{Discussion and outlook}
With this first demonstration of laser-based CEMS, an entirely new class of materials is now compatible with laser nuclear spectroscopy. 
By implanting \thor{} into low band gap materials and observing both the lifetime of the internal conversion decay channel and the isomer shift of the transition, detailed information can be gathered about the local phononic, electronic, and nuclear structure. 
This in turn allows the isomeric transition to serve as a sensor of strain and temperature in the solid~\cite{Rellergert2010,Higgins2025}.
Further, at the energy scale of the IC electrons, their inelastic mean free path is a sensitive function of their energy relative to the Fermi level~\cite{seah_dench_imfp_electrons}. 
In the future, it may be possible to relate the efficiency with which IC electrons emerge from a surface with both the surface quality and the band structure local to the implanted \thor{}. 
Looking further forward, off-resonant excitation of \thor{} nuclei through a combined nuclear and phononic transition in the solid could enable laser-based nuclear resonance vibrational spectroscopy~\cite{lin2011_nrvs_fe2o3,nrvs_review_2021}. 
This could enable measurement of the local phonon density of states with sub-$\mu$eV precision, considerably beyond what is possible with conventional M\"{o}ssbauer spectroscopy~\cite{nrvs_review_2021}.

Given our previous work in \thf{} films~\cite{zhang2024thf}, it can be concluded that as the \thor{} is converted from the oxide to the fluoride state, the \thor{} isomer will convert from decaying by IC to decaying by the emission of a VUV photon. 
Thus, the relative ratio of radiative decay to conversion electron decay following laser excitation could provide a new battery of methods to characterize the chemical state of Th compounds, perhaps aiding studies of nuclear power generation~\cite{banerjee2016_th_nuclear_power}.

Additionally, laser-based CEMS provides a new platform to realize a solid-state Th clock that benefits from the inherent ease of a production of stoichiometric Th compounds, a $10^8$ reduction in clock interrogation cycle, and the potential for a current-based readout allowing simplification and miniaturization of future nuclear clocks.

\section{Acknowledgement}
This work was supported by NSF awards PHYS-2013011, PHY-2412869 and PHY-2207546, and ARO award W911NF-11-1-0369. PGT and KS acknowledge support by BaCaTeC (Grant 7 [2029-2]). 
ERH acknowledges institutional support by the NSF QLCI Award OMA-2016245.

This work used Bridges-2 at Pittsburgh Supercomputing Center through allocation PHY230110 from the Advanced Cyberinfrastructure Coordination Ecosystem: Services \& Support (ACCESS) program, which is supported by National Science Foundation grants \#2138259, \#2138286, \#2138307, \#2137603, and \#2138296.
DAR acknowledges support from the Institutional Computing Program at Los Alamos National Laboratory, via the Center for Integrated Nanotechnologies, a DOE BES user facility, for computational resources.

\bibliography{ref,Th229-apd,ThoriumSearch,tho2_properties,HM_all_refs_fixed,library-apd_fixed,Udeshika}
%SI/Harry_refs
\newpage

\include{SI.tex}

\end{document}

%% file: SI.tex
\section{Supplemental Information}
\section{Expected Internal Conversion Signal}

The incident VUV flux 
is attenuated, so that it follows $\varphi_0 e^{-\alpha z}$ where $\alpha$ is the VUV attenuation coefficient.
The number of excited thorium atoms in a target of length $L$ should go as~\cite{vonderWense2020_theory_of_direct_excitation}
\begin{equation}
\begin{aligned}
	N_e & 
    \approx \frac{4}{6}\frac{\lambda^2}{2 \pi} \frac{n_\text{Th}}{\Gamma_L }\frac{T|\tilde{n}|^2}{\text{Re}[\tilde{n}]} \frac{\varphi_0(1-e^{-\alpha L})}{\alpha} \times \frac{1}{1 + 4\left(\frac{\delta}{\Gamma_L}\right)^2} \times \left( \frac{t_e}{\tau_\text{rad}} \right),
\end{aligned}    
\end{equation} where $\lambda$ is the vacuum transition wavelength, $n_\text{Th}$ is the density of \thor{} in the \thox{} target, $\Gamma_\text{rad} = 1/\tau_\text{rad}$ is the vacuum radiative decay rate, $\Gamma_L$ is the VUV laser bandwidth, $\Gamma_\text{IC} = 1/\tau_\text{IC}$ is the IC decay rate, $\delta$ is the laser detuning, $\varphi_0$ is incident the laser photon flux, $L$ is the target thickness, $T$ is the transmission of the VUV laser into the target, and $\tilde{n} = n-i\kappa$ is the complex index of refraction with $\kappa = \lambda\alpha/4\pi$. The target was produced using a 0.75:0.25 \thor{}:$^{232}$Th isotope mix from Oak Ridge National Lab, leading to an effective \thor{} density of $n_\text{Th} = 0.75\times2.28\times10^{22}$ cm$^{-3}$~\cite{belle_berman_1984_tho2_report}.

\begin{table}[h!]
\caption{Parameter values and estimated relative errors used in calculation of expected IC signal. Relative uncertainties labeled by  an asterisk (*) follow log-normal distribution and should be read as $\ln(X/\bar{X})$. 0 error entries correspond to assumed values from literature.}
\begin{ruledtabular}
 \begin{tabular}{ccc} 
 Parameter & Value & Rel. Err.\\ [0.5ex] 
 \hline 
 $n_\text{Th}$ & $1.71\times10^{22}$ cm$^{-3}$ & 0 (Ref.~\cite{belle_berman_1984_tho2_report}) \\
 $L$ & 10 nm & 0.7*  \\ 
 $\Gamma_L$ & $2\pi \times 12$ GHz & 0.25 \\
 $\alpha$ & 0.1 nm$^{-1}$ & 0 (Ref.~\cite{tho2_abs_ellipsometry}) \\
 $n$ & 2.34 &  0 (Ref.~\cite{tho2_abs_ellipsometry}) \\
 $t_i$ & 6 $\mu$s & 1e-4 \\
 $\tau_\text{rad}$ & 1800 s & 0.07\\
 $\tau_\text{IC}$ & 10 $\mu$s & 0.1\\
 $T\eta_\text{e}$ & $1\times10^{-4}$ & 0.7* \\
 $\eta_\text{c}$ & 0.7 & 0.35* \\
\end{tabular}
\label{tab:ic_values}   
\end{ruledtabular}
\end{table}

From the number of excited \thor{} nuclei, the number of detected IC electrons is given by
\begin{equation}
    N_\text{det} = \eta_\text{e}\eta_\text{c} N_e \times e^{-t_i/\tau_\text{IC}}
\end{equation} where $\eta_\text{e}$ is the extraction efficiency from the \thox{} target, $\eta_\text{c}$ is the collection efficiency, and $t_i$ is the start of the acquisition time window. The extraction efficiency represents the probability that an IC electron is able to leave the \thox{} target, and combines many physical processes such as whether the electron is promoted high enough into the conduction band to overcome the work function barrier, whether the electron inelastically scatters, whether surface conditions are favorable, etc. 
Due to all of these confounding factors, theoretical calculation of $\eta_\text{e}$ is difficult, and it must be estimated from experiment. 
To do this, we make the assumption that the probability for a photoelectron promoted by a VUV photon to leave the target is the same as an IC electron promoted by an excited \thor{} nucleus. By making this assumption we are able to use the efficiency with which photoelectrons are generated by our VUV laser system to obtain an estimate of $T\eta_\text{e}$. The collection efficiency $\eta_\text{c}$ is simply the efficiency with which electrons that leave the target are detected on the MCP given the voltage biasing and magnetic field configuration used in the experiment.
This can readily be determined by measuring the ratio of the number of photoelectrons collected on the MCP front plate relative to the number of photoelectrons leaving the target.

Using the values listed in Table~\ref{tab:ic_values}, the expected on-resonance IC signal was estimated to be $N_\text{det}/(\hbar \omega_0\varphi_0t_\text{e}) =  0.15_{-0.09}^{+0.25}$ $e^-$/$\mu$J/shot.

\textbf{Quenching of ASE in Xe Four-wave Mixing} \\
The four-wave mixing process can produce a beam of amplified spontaneous emission (ASE) along the beam axis, complicating detection.
The origin of the ASE is resonant excitation by the 249.6~nm laser of the $5p^{6 ~1}S_0 ~\rightarrow$ $~5p^5\left(^2P^{\circ}_{3/2} \right) 6p~^2\left[1/2\right]_0$ transition.
Any Xe population not participating in the four-wave mixing process will be left in the excited two-photon state. 
This $5p^5\left(^2P^{\circ}_{3/2} \right) 6p~^2\left[1/2\right]_0$ (abbreviated as $6p~^2\left[1/2\right]_0$) state will then decay to the $5p^5\left(^2P^{\circ}_{3/2} \right) 6s~^2\left[3/2\right]^{\circ}_1$ (abbreviated as $6s~^2\left[3/2\right]^{\circ}_1$) by 828 nnm emission in $\approx$30 ns, and the $6s~^2\left[3/2\right]^{\circ}_1$ state will decay to the $^1S_0$ ground state by emission of a 147 nm photon in $\approx$3.7 ns. 
If the Xe pressure is in the several hundred Pa range, as it is for efficient four-wave mixing, there are enough nearby Xe atoms that they may re-absorb these 828 nm and 147 nm photons. 
Effectively, this leads to radiation trapping which extends the effective fluorescence lifetime of the Xe spontaneous emission~\cite{alekseev_xe_ase_quenching}. 
At the same time the 828/147 nm spontaneous emission will experience gain as it stimulates Xe in the excited $6p~^2\left[1/2\right]_0$ and $6s~^2\left[3/2\right]^{\circ}_1$ to emit, yielding amplified spontaneous emission (ASE). 
This gain will be highly directional, since the excited Xe will essentially lie in a column defined by the propagation of the 249.6 nm pulse of the pump laser. 
This interplay between radiation trapping and ASE will then yield bi-directional emission from the Xe cell along the pumping axis, which will have a timescale much longer than the spontaneous emission lifetime of the excited states~\cite{RANKIN_bidirectional_ase}.

To mitigate this effect, we introduce \ce{N2} gas in the Xe, which quenches the excited state Xe population via collision.

\begin{figure}
    \centering
    \includegraphics[width=0.9\linewidth]{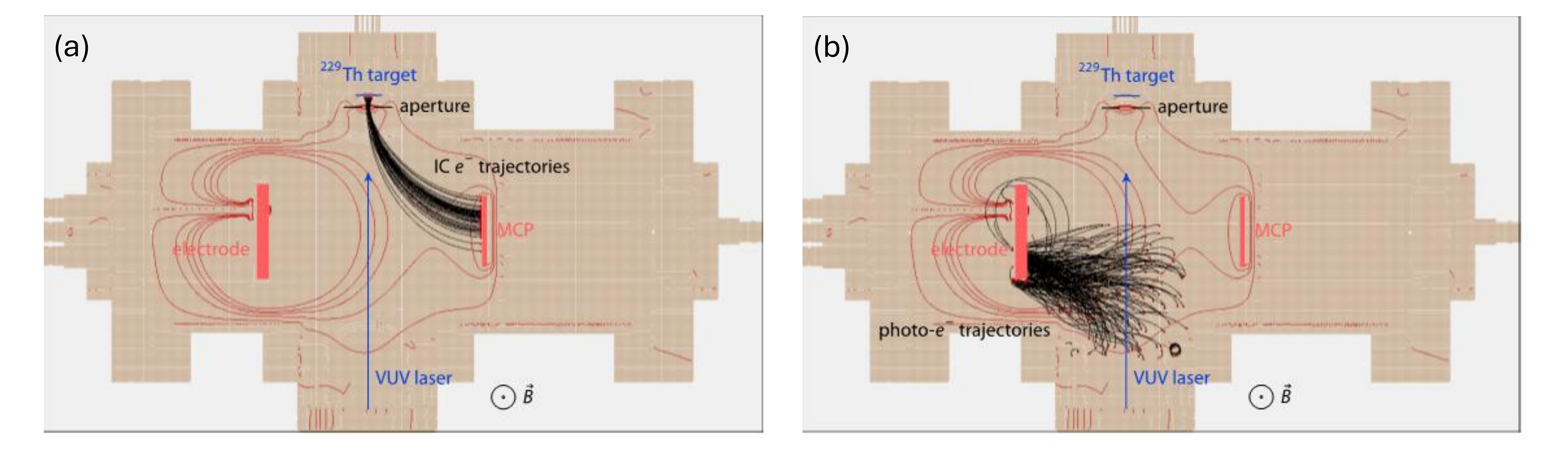}
    \caption{SIMION simulations of electron trajectories under the voltage biasing and static magnetic field used in the experiment. (a) Trajectories of IC electrons from the target region. The magnetic field bends them towards the detection MCP through a saddle in the electric potential. (b) The trajectories of photoelectrons generated by VUV scattered light at random locations in the chamber. Most either crash into the chamber walls or end up falling into the electrode biased at +2500V.}
    \label{fig:simion}
\end{figure}

\textbf{Simulation of Electron Trajectories}\\
SIMION simulations were carried out in order to determine a combination of voltage biasing and static magnetic field that would guide IC electrons to our detector while diverting background photoelectrons to a secondary electrode. The voltage biasing chosen during the IC observation period was to have the target at -405 V, the front aperture of the target mount at +135V, the front of the detection MCP at +140 V, and the secondary electrode at +2500V. The magnetic field was set to $\approx$3-5 G. As can be seen in Fig.~\ref{fig:simion}(a), under the voltage biasing described there is a "saddle" in the potential through which the IC electrons that have been accelerated by our target mount system can be made to travel through by the magnetic field. Meanwhile, as can be seen in Fig.~\ref{fig:simion}(b), electrons that are generated randomly throughout the chamber with $\sim 8$ eV of kinetic energy either crash into the chamber wall or fall into the sacrificial electrode.

\section{Preparation of the Target}
The target used in this study was prepared in an electrodeposition buffer solution of NaOH and sulfuric acid. \thor{} material was deposited onto the stainless steel disk until target activity was reached, whereupon a quench was performed with ammonium hydroxide. The disk was then heated overnight at 200-300 $^{\circ}$C in ambient atmosphere.

While the electrodeposited target is likely a mixture of Th hydroxides, oxides, and metal impurities, the use of low impurity ($>$99.99\% grade) chemicals during electrodeposition and heat treatment in air lead us to assume it is predominantly Th in an oxidized state. As the most stable oxide form, we assume the target is predominately \thox{}.
\section{Internal Conversion Rate derivation}
\label{SI:Sec:ICRate}

\begin{figure}[htb]
  \centering
\begin{tikzpicture}[line width=1pt,scale=1]
  % smaller separation for double lines
  \def\sep{0.5pt}
  % key points
  \coordinate (vL) at (1,0.5);
  \coordinate (vR) at (3,0.5);
  \coordinate (A1) at (0.4,1.4);
  \coordinate (A2) at (0.4,-0.4);
  \coordinate (B1) at (3.6,1.4);
  \coordinate (B2) at (3.6,-0.4);

  % --- Black single incoming lines with mid arrows ---
  \draw[black, line width=0.6pt] (A1) -- (vL);
  \draw[black, line width=0, mid arrow ]  (vL) --(A1) ;
  \draw[black, line width=0.6pt] (A2) -- (vL);
  \draw[black, line width=0, mid arrow] (A2) -- (vL);

 % Virtual interaction (red snake) with black W label
  \draw[red, decoration={snake,amplitude=1.5pt,segment length=6pt},decorate]
    (vL) -- node[above,yshift=2pt,font=\small,text=black]{\(W\)} (vR);

  % --- Blue double outgoing lines with true parallel strokes ---
  % upper stroke
  \draw[blue, line width=0.6pt] ($(vR)!\sep!90:(B1)$) -- ($(B1)!\sep!-90:(vR)$);
  % lower stroke
  \draw[blue, line width=0.6pt] ($(vR)!\sep!-90:(B1)$) -- ($(B1)!\sep!90:(vR)$);
  % mid arrow on upper leg
  \draw[blue, line width=0, mid arrow]  (vR) --  (B1) ;

  % lower leg strokes
  \draw[blue, line width=0.6pt] ($(vR)!\sep!90:(B2)$) -- ($(B2)!\sep!-90:(vR)$);
  \draw[blue, line width=0.6pt] ($(vR)!\sep!-90:(B2)$) -- ($(B2)!\sep!90:(vR)$);
  \draw[blue, line width=0, mid arrow] (B2)--(vR) ;

  % --- State labels ---
  \node[font=\small] at (3.9,1.4){\(\lvert g\rangle\)};
  \node[font=\small] at (3.9,-0.4){\(\lvert e\rangle\)};
  \node[font=\small] at (-0.2 ,1.4){\(\lvert c \mathbf{k} \sigma_{\!c}\rangle\)};
  \node[font=\small] at (-0.2,-0.4){\(\lvert v \mathbf{k} \sigma_{\!v}\rangle\)};

\end{tikzpicture}
  \caption{%
 Feynman diagram of the internal conversion process. Single black strokes represent the electron lines, while the double blue ones denote the nuclear lines. The interaction is mediated by a virtual photon exchange, corresponding to the hyperfine interaction $W$. The crystal momentum $\mathbf{k}$ is conserved in this process. 
   }
  \label{fig:internal-conversion}
\end{figure}
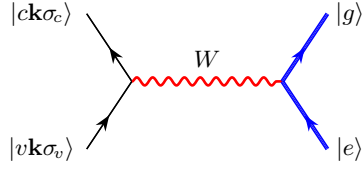

We consider the IC processes in an insulator, when the bandgap $\Delta$ is smaller than the nuclear isomer energy $\omega_\text{nuc}$. In the IC process,
the nuclear excitation is transferred to the electrons, spawning a particle-hole pair, with a valence band electron promoted into the conduction band leaving behind a hole in the valence band, see Fig.~\ref{fig:internal-conversion}.

The \thor{} nuclear subsystem is modeled as two distinct energy levels: the ground state $|g\rangle$ with nuclear spin $I_g = 5/2$ and the excited (isomeric) state $|e\rangle$ with $I_e = 3/2$, separated by the energy gap $\omega_{\mathrm{nuc}}$.

The IC process is mediated by the hyperfine interaction (HFI). In a crystal, containing $\mathcal{N}$ \thor{} nuclei with one \thor{} per unit cell, HFI reads
\begin{align}
    W(\mathbf{r}) = \sum_{\nu=1}^\mathcal{N} \mathcal{M}(\mathbf{R}_\nu) \cdot \mathcal{T}(\mathbf{r} - \mathbf{R}_\nu) \,, \label{Eq:WHFI}
\end{align}
where we sum over unit cells, $\mathcal{M}(\mathbf{R}_\nu)$ is a \thor{} nuclear magnetic moment operator and the \thor{}-centered $\mathcal{T}$  is a rank-1 tensor acting on the electronic degrees of freedom. The nuclear electric-quadrupolar and higher-rank HFI interactions can be added in a similar fashion. For \thox{}, due to the cubic symmetry, the electric-quadrupole contribution vanishes.

To compute the IC rate, we use the Fermi's golden rule.
The initial state of the electron subsystem is a fully-occupied valence band $\ket{\Tilde{0}}$, and the final state is the particle-hole excitation $a_{c\mathbf{k} \sigma_c}^\dagger a_{v\mathbf{q} \sigma_v} \ket{\Tilde{0}}$ with energy $ \varepsilon_{c\mathbf{k}} - \varepsilon_{v\mathbf{q}}$, where $\varepsilon_{c\mathbf{k}}$ and $\varepsilon_{v\mathbf{q}}$ are the band functions of the conduction (c) and the valence (v) bands.  Then, 
the final state quantum numbers are spanned by crystal momenta $\mathbf{k}$, $\mathbf{q}$ and two electron spin projections. We also sum over the nuclear  ground state magnetic quantum numbers $M_g$ and average over the nuclear isomer state magnetic quantum numbers $M_e$,
\begin{align}
 \Gamma_\text{IC}^{(\mathcal{N})}  =  \frac{2 \pi}{\hbar} \frac{1}{2I_e+1} \sum_{M_gM_e} \sum_{\sigma_c\sigma_v} 
 \int_\text{BZ}  d^3 k \, \int_\text{BZ}  d^3 q |\mel{c \mathbf{k} \sigma_c }{W^{ge}}{ v \mathbf{q}\sigma_v }|^2
  \delta(\varepsilon_{c\mathbf{k}} - \varepsilon_{v\mathbf{q}}-\omega_\text{nuc}) \, . \label{Eq:ICRateStart}
\end{align}
The integrations are carried over the  Brillouin zone (BZ). We use conventions of Ref.~\cite{callaway1991Book} in our derivation.

The HFI, Eq.~\eqref{Eq:WHFI} is cell-periodic, and its matrix element
$\mel{c \mathbf{k} \sigma_c }{W^{ge}}{ v \mathbf{q}\sigma_v }$
between Bloch functions can be reduced from an integration over the entire crystal to an an integration over a single unit cell. The result reads
\begin{align}
\mel{c\mathbf{k} \sigma_c}{W^{ge}}{v\mathbf{q} \sigma_v} =
 \delta( \mathbf{k}-\mathbf{q} ) W^{ge}_{c\sigma_c v \sigma_v}(\mathbf{k}) 
\end{align} 
with 
\begin{align}\label{eq:W_mel_nonrel}
W^{ge}_{c\sigma_cv \sigma_v}(\mathbf{k}) 
\equiv  \frac{(2 \pi)^3}{\Omega} \mel{g}{\mathcal{M}}{e} \cdot \int_{\Omega} d^3r \, 
u^*_{c\mathbf{k}}(\mathbf{r}) \chi^\dagger_{\sigma_c} \mathcal{T}(\mathbf{r})  \chi_{\sigma_v}  u_{v\mathbf{k}}(\mathbf{r})  \,.
\end{align}
Here $\Omega$ is the unit cell volume, $u_{\ldots}(\mathbf{r})$ are cell-periodic envelopes of Bloch functions~\cite{callaway1991Book}, and $\chi$ are the conventional electron spinors. Note that the cell-periodicity imposes conservation of crystal momentum, $\mathbf{k} = \mathbf{q}$.

Now with this matrix element, we evaluate the IC rate.
\begin{align}
 \Gamma_\text{IC}^{(\mathcal{N})}  = 
  \frac{2 \pi}{\hbar} 
  \frac{1}{2I_e+1} \sum_{M_gM_e} \sum_{\sigma_c\sigma_v} 
 \int d^3 k \, \int  d^3 q 
 |W^{ge}_{c\sigma_c v \sigma_v}(\mathbf{k}) |^2
 \delta( \mathbf{k}-\mathbf{q} ) \delta( \mathbf{k}-\mathbf{q} )
  \delta(\varepsilon_{c\mathbf{k}} - \varepsilon_{v\mathbf{q}}-\omega_\text{nuc})
\end{align}
There is a product of two identical delta functions.  $\mathbf{q}$ is replaced by $\mathbf{k}$ while integrating over $\mathbf{q}$,  but we encounter $\delta(\mathbf{0}) =\lim_{\Delta\mathbf{q} \rightarrow 0} \delta( \Delta\mathbf{q} )$.
Using the following identity for each component of $\Delta\mathbf{q}$:
$$
\lim_{\Delta q_x \rightarrow 0} \delta( \Delta q_x ) = 
\lim_{L_x \rightarrow \infty} \lim_{\Delta q_x \rightarrow 0} \frac{1}{2\pi} \int_{-L_x/2}^{L_x/2} e^{i \Delta q_x x} dx = \frac{ L_x }{2\pi} \,,
$$
where $L_x$ is the crystal size in the x-direction, we can show that
$$
\lim_{\Delta\mathbf{q} \rightarrow 0} \delta( \Delta\mathbf{q} )  = \frac{ V_\text{xtal} }{(2\pi)^3} \,,
$$ 
where the crystal volume $V_\text{xtal}=L_x L_y L_z$.
This is similar to the formal time-domain limit in deriving the Fermi's golden rule, see, e.g., p.~72 of Ref.~\cite{bethe1955mesons-book}.

Thereby, the IC rate per  crystal volume
\begin{align}\label{eq:IC_rate}
 \Gamma_\text{IC}^{(\mathcal{N})}/V_\text{xtal} =  
  \frac{2 \pi}{\hbar} \frac{1}{(2 \pi)^3}
  \frac{1}{2I_e+1} \sum_{M_gM_e} \sum_{\sigma_c\sigma_v} 
 \int_\text{BZ} d^3 k \, 
|W^{ge}_{c\sigma_c v \sigma_v}(\mathbf{k}) |^2
  \delta(\varepsilon_{c\mathbf{k}} - \varepsilon_{v\mathbf{k}}- \hbar \omega_\text{nuc})
\end{align}

% One can compare this derived IC rate to the absorption rate~\cite{callaway1991Book} of electromagnetic wave of  frequency $\omega_\text{nuc}$, which also leads to a rate per crystal volume. 

The assumption made during the derivation was that all $\mathcal{N}$ \thor{} nuclei were initially in the excited (isomer) state. Consider an ensemble of quantum emitters, with $N(t)$ being the number of emitters in the excited state at time $t$. With a single emitter decay rate $\gamma$ and 
\begin{align}
    dN/dt &= - \gamma N(t) \Rightarrow N(t) = N(0) \exp{-\gamma t}  \nonumber\\
    &\Rightarrow\text{Ensemble decay rate $\Gamma$} = -dN/dt = \gamma  N(0) \exp{-\gamma t} \,.
\end{align}
We see that the ensemble decay rate $\Gamma$ at $t=0$ is  $\gamma  N(0)$. The experimentally relevant excited state population, however, decays as $\exp{-\gamma t}$. Based on this discussion, we define the experimentally measured IC decay rate
\begin{align}
    \Gamma_\text{IC}= \Gamma^{(\mathcal{N})}_\text{IC}/\mathcal{N} = 
    \Omega \Gamma_\text{IC}^{(\mathcal{N})}/V_\text{xtal} \, .
\end{align}
Following the  derivation of interband electromagnetic absorption rates~\cite{callaway1991Book}, we define a surface $S_E$ of constant energy in  $\mathbf{k}$-space through an implicit relation $\varepsilon_{c\mathbf{k}} - \varepsilon_{v\mathbf{k}} =  \hbar \omega_\text{nuc}$.  If the matrix element remains reasonably constant on $S_E$, the rate simplifies to 
\begin{align}
\boxed{
 \Gamma_\text{IC} =  \frac{1}{\tau_\text{IC}} =
  \frac{2 \pi}{\hbar }  \,  
  \frac{1}{2I_e+1} \sum_{M_gM_e} \sum_{\sigma_c\sigma_v} 
\overline{|{W^{ge}_{c\sigma_c v \sigma_v}(\mathbf{k})}  |^2} G_{cv}(\hbar\omega_\text{nuc}) \,,
}
\label{Eq:gammaIC-final}
 \end{align}
with the joint density of states (JDOS)
\begin{align}\label{Eq:JDOS}  
 G_{cv}(E) =  
\frac{\Omega}{(2 \pi)^3} \int_\text{BZ} d^3 k \, \delta(\varepsilon_{c\mathbf{k}} - \varepsilon_{v\mathbf{k}}- E) =\frac{\Omega}{(2 \pi)^3} \int \frac{dS_E}{
|\nabla_\mathbf{k}(\varepsilon_{c\mathbf{k}} - \varepsilon_{v\mathbf{k}})_{E}|} \,.
\end{align}
With this simplification, $\overline{|{W^{ge}_{c\sigma_c v \sigma_v}(\mathbf{k})}  |^2}$ has the meaning of $|{W^{ge}_{c\sigma_c v \sigma_v}(\mathbf{k})}  |^2$  averaged over the surface of constant energy $S_E$.

This concludes the derivation of the IC rate, Eq.(1) of the main text. 

As a side note, notice that the electromagnetic absorption coefficient at laser frequency $\omega_\text{nuc}$ is proportional to the same JDOS $G_{cv}(\hbar\omega_\text{nuc})$. This can be used to back out the JDOS value from the laser absorption measurements at a frequency slightly detuned away from $\omega_\text{nuc}$.

To proceed with the computation of the hyperfine matrix element, we expand the functions $u*_{c{\bf k}}({\bf r})$ in terms of the Th atomic states. Due to the short-range nature of the hyperfine interaction, relativistic effects play an important role. As a result, we use relativistic Dirac spinors for the Th atomic states. However, Eq.~\eqref{eq:W_mel_nonrel} is given in terms of nonrelativistic (or scalar relativistic~\cite{koelling1977technique}) two-component spinors. Here, for simplicity, we replace Eq.~\eqref{eq:W_mel_nonrel} with
\begin{align}\label{eq:W_mel_rel}
W^{ge}_{c\sigma_cv \sigma_v}(\mathbf{k}) 
\equiv  \frac{(2 \pi)^3}{\Omega} \mel{g}{\mathcal{M}}{e} \cdot \int_{\Omega} d^3r \, 
u^\dagger_{c\mathbf{k}\sigma_c}(\mathbf{r}) \mathcal{T}(\mathbf{r})  u_{v\mathbf{k}\sigma_v}(\mathbf{r})  \,,
\end{align}
where $u_{n\mathbf{k}\sigma_n}(\mathbf{r})$ is now a four-component Dirac spinor with momentum $\bf k$ and spin projection $\sigma_n$.

The crystal function $u_{c\mathbf{k}\sigma_c}(\mathbf{r})$ (and similarly $u_{v\mathbf{k}\sigma_v}(\mathbf{r})$)  may be expanded in terms of the Th atomic states of definite principle and angular momentum quantum numbers $\varphi_{njlm}({\bf r})$
\begin{equation}\label{eq:expansion_Th}
    u_{c{\bf k}\sigma_n}({\bf r})=\sqrt{\frac{\Omega}{(2\pi)^3}}\sum_{njlm} a_{njlm}(c{\bf k}\sigma_n)\varphi_{njlm}({\bf r})+\ldots\,,
\end{equation}
where the ellipses denote contributions from O atomic states, which, due to their insignificant overlap with the Th nucleus, do not affect the HFI matrix element. Using the expansion~\eqref{eq:expansion_Th}, we may write
\begin{align}\label{eq:HFI_mel_expanded}
W^{ge}_{c\sigma_cv\sigma_v}(\mathbf{k}) = \sum_{njlm}\sum_{n'j'l'm'} a^*_{njlm}(c{\bf k}\sigma_c)a_{n'j'l'm'}(v{\bf k}\sigma_v) \tilde{W}^{ge}_{njlmnj'l'm'}\,,
\end{align}
where $\tilde{W}^{ge}_{njlmnj'l'm'}=\mel{g}{\mathcal{M}}{e} \cdot\int_{\Omega} d^3r \varphi^\dagger_{njlm}(\mathbf{r})\mathcal{T}(\mathbf{r})\varphi_{n'j'l'm'}(\mathbf{r})$.

Given the crystal functions and the atomic wavefunctions, the expansion coefficients $a_{njlm}({\bf k}\sigma_n)$ may be computed and the total HFI matrix element obtained. Here, we make a simplifying assumption that only a few terms contribute significantly to the sum~\eqref{eq:HFI_mel_expanded}. Furthermore, we shall neglect the contribution cross terms in $\left|W^{ge}_{c\sigma_cv\sigma_v}(\mathbf{k})\right|^2$. With these simplifications, the IC rate may be written as
\begin{align}\label{eq:IC_rate_mod_1_SI}
 \Gamma_\text{IC}&\approx  
  \frac{\pi}{\hbar}
  \frac{1}{2I_e+1} \sum_{M_gM_e}  \sum_{njlm}\sum_{n'j'l'm'}\left|\tilde{W}^{ge}_{njlmn'j'l'm'}\right|^2\nonumber\\
  &\times\sum_{\sigma_c\sigma_v}
 \frac{2\Omega}{(2 \pi)^3} \int_\text{BZ} d^3 k\left|a_{njlm}(c{\bf k}\sigma_c)\right|^2\left|a_{n'j'l'm'}(v{\bf k}\sigma_v)\right|^2
  \delta(\varepsilon_{c\mathbf{k}} - \varepsilon_{v\mathbf{k}}- \hbar \omega_\text{nuc})\,.
\end{align}

Let us now consider the matrix element $W^{ge}_{c\sigma_cv\sigma_v}(\mathbf{k})$ in more details. Using the Wigner-Eckart theorem, we may write
\begin{align}
    \tilde{W}^{ge}_{njlmnj'l'm'}=\sum_{\nu=-1}^1(-1)^{\nu+I_g-M_g+j-m}\begin{pmatrix}
    I_g & 1 & I_e \\
    -M_g & \nu & M_e
\end{pmatrix}\begin{pmatrix}
    j & 1 & j' \\
    -m& -\nu & m'
\end{pmatrix}\langle g||\mathcal{M}||e\rangle\langle njl||\mathcal{T}||n'j'l'\rangle\,.
\end{align}
Here, $\langle g||\mathcal{M}||e\rangle\approx0.84\mu_N$ ($\mu_N$ is the nuclear magneton) is the reduced matrix element of the nuclear magnetic moment operator (see, e.g.~Ref.~\cite{morgan_internal_conversion_2024}), and $\langle njl||\mathcal{T}||n'j'l'\rangle$ is the reduced matrix element of the rank-1 electronic HFI tensor. 
Th off-diagonal HFI matrix elements are at least two orders of magnitude smaller than the diagonal ones, so the most significant contributions to the IC rate come from terms with $(njl)=(n'j'l')$. The reduced matrix element $\langle njl||\mathcal{T}||njl\rangle$ may be related to the ground state HFI constant $A_{njl}$ via~\cite{johnson2007atomic}
\begin{equation}
    A_{njl}=\frac{\mu_g}{I_gj}\frac{(2j)!}{\sqrt{(2j-1)!(2j+2)!}}\langle njl||\mathcal{T}||njl\rangle\,,
\end{equation}
where $\mu_g = 0.360(7) \mu_N$ is the  magnetic moment of the ground nuclear state~\cite{SafSafRad2013-Th3plus}.
With this, the summations over $M_g$ and $M_e$ in Eq.~\eqref{eq:IC_rate_mod_1_SI} may be carried out, giving
\begin{equation}
    \sum_{M_gM_e}\left|\tilde{W}^{ge}_{njlmn'j'l'm'}\right|^2=\frac{1}{3}\left[\sum_{\nu=-1}^1\begin{pmatrix}
    j & 1 & j' \\
    -m& \nu & m'
\end{pmatrix}^2\right]\langle g||\mathcal{M}||e\rangle^2\langle njl||\mathcal{T}||n'j'l'\rangle^2\,.
\end{equation}

To proceed, we make a further assumption that the quantity in the second line of Eq.~\eqref{eq:IC_rate_mod_1_SI} does not depend strongly on the magnetic quantum numbers $m$ and $m'$, i.e., 
\begin{align}
    &\sum_{\sigma_c\sigma_v}
 \frac{2\Omega}{(2 \pi)^3} \int_\text{BZ} d^3 k\left|a_{njlm}(c{\bf k}\sigma_c)\right|^2\left|a_{n'j'l'm'}(v{\bf k}\sigma_v)\right|^2
  \delta(\varepsilon_{c\mathbf{k}} - \varepsilon_{v\mathbf{k}}- \hbar \omega_\text{nuc})\nonumber\\
  &\approx\sum_{\sigma_c\sigma_v}
 \frac{2\Omega}{(2 \pi)^3} \int_\text{BZ} d^3 k\left|a_{njl}(c{\bf k}\sigma_c)\right|^2\left|a_{n'j'l'}(v{\bf k}\sigma_v)\right|^2
  \delta(\varepsilon_{c\mathbf{k}} - \varepsilon_{v\mathbf{k}}- \hbar \omega_\text{nuc})\,,
\end{align}
then the summation over $m$ and $m'$ may be carried out analytically, giving finally
\begin{align}\label{eq:IC_rate_mod_1_SI_summed}
 \Gamma_\text{IC}&\approx  
  \frac{\pi}{3\hbar}
  \frac{1}{2I_e+1} \langle g||\mathcal{M}||e\rangle^2 \sum_{njl}\sum_{n'j'l'}\langle njl||\mathcal{T}||n'j'l'\rangle^2\nonumber\\
  &\times\sum_{\sigma_c\sigma_v}
 \left[\frac{2\Omega}{(2 \pi)^3} \int_\text{BZ} d^3 k\left|a_{njl}(c{\bf k}\sigma_c)\right|^2\left|a_{n'j'l'}(v{\bf k}\sigma_v)\right|^2
  \delta(\varepsilon_{c\mathbf{k}} - \varepsilon_{v\mathbf{k}}- \hbar \omega_\text{nuc})\right]\,.
\end{align}

One recognizes the quantity in the square bracket as the (twice) projected joint density of states (PJDOS), which counts the number of allowed transitions at energy $\hbar\omega_{\rm nuc}$ between the atomic component $\ket{n'j'l'm'}$ of $\ket{v \mathbf{k}\sigma_v}$ and the atomic component $\ket{njlm}$ of $\ket{c \mathbf{k}\sigma_c}$. In Fig.~\ref{fig:ThO2 structure and PDOS}(c) in the main text, we present the most relevant PJDOS. These PJDOS were computed using VASP (see the next section) and are use the scalar relativistic approximation of Koelling \& Harmon~\cite{koelling1977technique}. To obtain the relativistic PJDOS, we adopt the procedure developed in Ref.~\cite{morgan_internal_conversion_2024}, i.e., for any $l>0$, $\left|a_{nl-1/2l}(c{\bf k}\sigma_c)\right|^2=\left|a_{nl+1/2l}(c{\bf k}\sigma_c)\right|^2=\left|a_{nl}\right|^2/2$ where $\left|a_{nl}\right|^2$ is the VASP projection. As a result, for $l>0$ the relativistic $(njl)\rightarrow(njl)$ PJDOS is a quarter of the scalar relativistic value. On the other hand, since the VASP calculation was spin-independent, the double summation over $\sigma_c$ and $\sigma_v$ in Eq.~\eqref{eq:IC_rate_mod_1_SI_summed} results in a factor of $2\times2=4$. 
In Table~\ref{tab:Th3+_hyperfine}, we list the HFI constants $A$, the corresponding ``diagonal'' values of the PJDOS at the nuclear energy for various Th atomic states, and the contibutions to the IC rate. Using these values, we arrive at an estimate for the IC rate of 
\begin{equation}
    \Gamma_{\rm IC}\approx 1.3\times10^4\,{\rm s}^{-1}\,,
\end{equation}
corresponding to an IC lifetime of 80 $\mu$s. 

\begin{table}[ht]
\caption{Magnetic‐dipole $A$ HFI constants for low‐lying levels of $^{229}$Th$^{3+}$, the corresponding values of PJDOS at the nuclear energy $\approx$ 8.35 eV, and the contributions to the IC rate. The $A$ constants for $7s$ and $6p$ states are computed using our relativistic atomic-structure code with random-phase-approximation (RPA) and perturbative Brueckner-orbital (BO) corrections. The $A$ constants for $6d$ and $5f$ states are taken from experiment~\cite{PhysRevLett.106.223001}.}
\label{tab:Th3+_hyperfine}
\begin{ruledtabular}
\begin{tabular}{cccc}
& $A$ (MHz) & PJDOS($\hbar\omega_{\rm nuc})$ (1/eV) & Contribution to $\Gamma_{\rm IC}$ ($s^{-1}$)\\
\hline
$7s_{1/2}-7s_{1/2}$ & $+6249$ & $5\times10^{-5}$ & 431\\
$6p_{1/2}-6p_{1/2}$ & $+8604$   &   $2.5\times10^{-4}$ & 8182\\
$6p_{3/2}-6p_{3/2}$ & $+426$     & $2.5\times10^{-4}$ & 200\\
$6d_{3/2}-6d_{3/2}$ & $+155.3$   &  $2.5\times10^{-3}$ & 267  \\
$6d_{5/2}-6d_{5/2}$ & $-12.6$     & $2.5\times10^{-3}$ & 6.14\\
$5f_{5/2}-5f_{5/2}$ & $+82.0$     & $2.5\times10^{-2}$ & 2601\\
$5f_{7/2}-5f_{7/2}$ & $+31.4$     & $2.5\times10^{-2}$ & 915\\
\end{tabular}
\end{ruledtabular}
\end{table}

\section{Computational methods for electronic structure theory}
\label{SI:Sec:Computational}

Calculations were performed with VASP~\cite{RN12}, version 6.4.2, using the PAW~\cite{RN14} method.
The structure of \ce{ThO2} was optimized with DFT in the conventional unit cell using the PBE~\cite{RN13} functional, 6-6-6 $\Gamma$-centred $k$-mesh and a 500 eV plane wave cutoff. 
Subsequent electronic structure calculations used the optimized structure in the primitive cell representation.

Parameters for $G_0W_0$ calculations~\cite{RN624,RN625,RN626}, specifically the $k$-mesh, plane wave cut-off energy, and the number of frequency grid points (NOMEGA tag in VASP), were tested for convergence of the band gap.
The number of unoccupied bands was 812 (there are 12 occupied bands).
Results of these tests are shown in Tables \ref{tab:G0W0 k-point convergence}, \ref{tab:G0W0 plane wave convergence}, and \ref{tab:G0W0 NOMEGA convergence}.
On the basis of these results, further $G_0W_0$ calculations were done with an 8-8-8 k-mesh, a 400 eV plane wave cut-off, and 80 frequency grid points.

\begin{table}[htb]
\caption{Convergence of the $G_0W_0$ band gap with respect to $k$-mesh.}
\begin{ruledtabular}
% \begin{tabular}{c|c|c}
\begin{tabular}{ccc}
mesh  & spacing (\AA{}$^{-1}$) & band gap (eV)  \\
\hline
6-6-6 & 0.05    & 6.31 \\
7-7-7 & 0.045   & 6.21 \\
8-8-8 & 0.04    & 6.18 \\
9-9-9 & 0.035   & 6.22
\end{tabular}
\label{tab:G0W0 k-point convergence}
\end{ruledtabular}
\end{table}

\begin{table}[htb]
\caption{Convergence of the $G_0W_0$ band gap with respect to plane wave cut-off.}
\begin{tabular}{cc}
\hline\hline
Cut-off (eV) & band gap (eV)   \\
\hline
400   & 6.31  \\
500   & 6.30  \\
600   & 6.29 \\
\hline\hline
\end{tabular}

\label{tab:G0W0 plane wave convergence}
\end{table}

\begin{table}[htb]
\caption{Convergence of the $G_0W_0$ band gap with respect to frequency grid points (NOMEGA).}
\begin{tabular}{cc}
\hline\hline
NOMEGA & band gap (eV) \\
\hline
8      & 7.45     \\
20     & 6.36     \\
40     & 6.21     \\
80     & 6.18    \\
\hline\hline
\end{tabular}
\label{tab:G0W0 NOMEGA convergence}
\end{table}

\begin{figure}[htb]
    \centering
    \includegraphics[width=\linewidth]{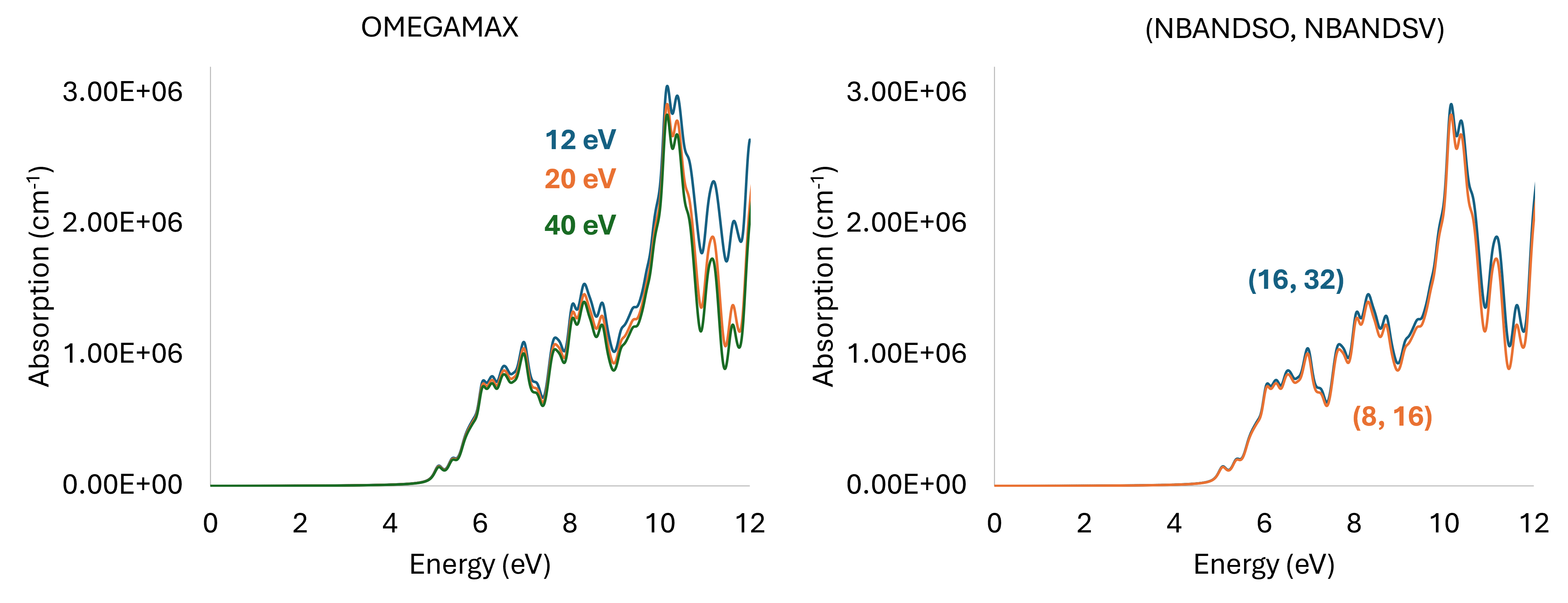}
    \caption{Absorption spectra of \ce{ThO2} computed with $G_0W_0$+BSE using different OMEGAMAX and (NBANDSO, NBANDSV) settings. The meanings of these parameters are given in the text.}
    \label{fig:BSE convergence absorption spectra}
\end{figure}

The Bethe-Salpeter equation (BSE) is a two-particle Green's-function formalism to explicitly account for electron–hole interactions in electronic excited states~\cite{Albrecht1998,Rohlfing1998}. The
$G_0W_0$+BSE parameters were converged with respect to the predicted absorption spectrum.
The tested parameters were the highest excitation energy (OMEGAMAX) and the number of occupied and unoccupied bands (NBANDSO and NBANDSV) considered in the BSE calculation.
Absorption spectra computed with various settings for these methods are shown in Fig.~\ref{fig:BSE convergence absorption spectra}.
On the basis of these tests, further $G_0W_0$+BSE calculations were done with OMEGAMAX = 20 eV and (NBANDSO, NBANDSV) = (8, 16).

\begin{figure}[htb]
    \centering
    \includegraphics[width=0.5\linewidth]{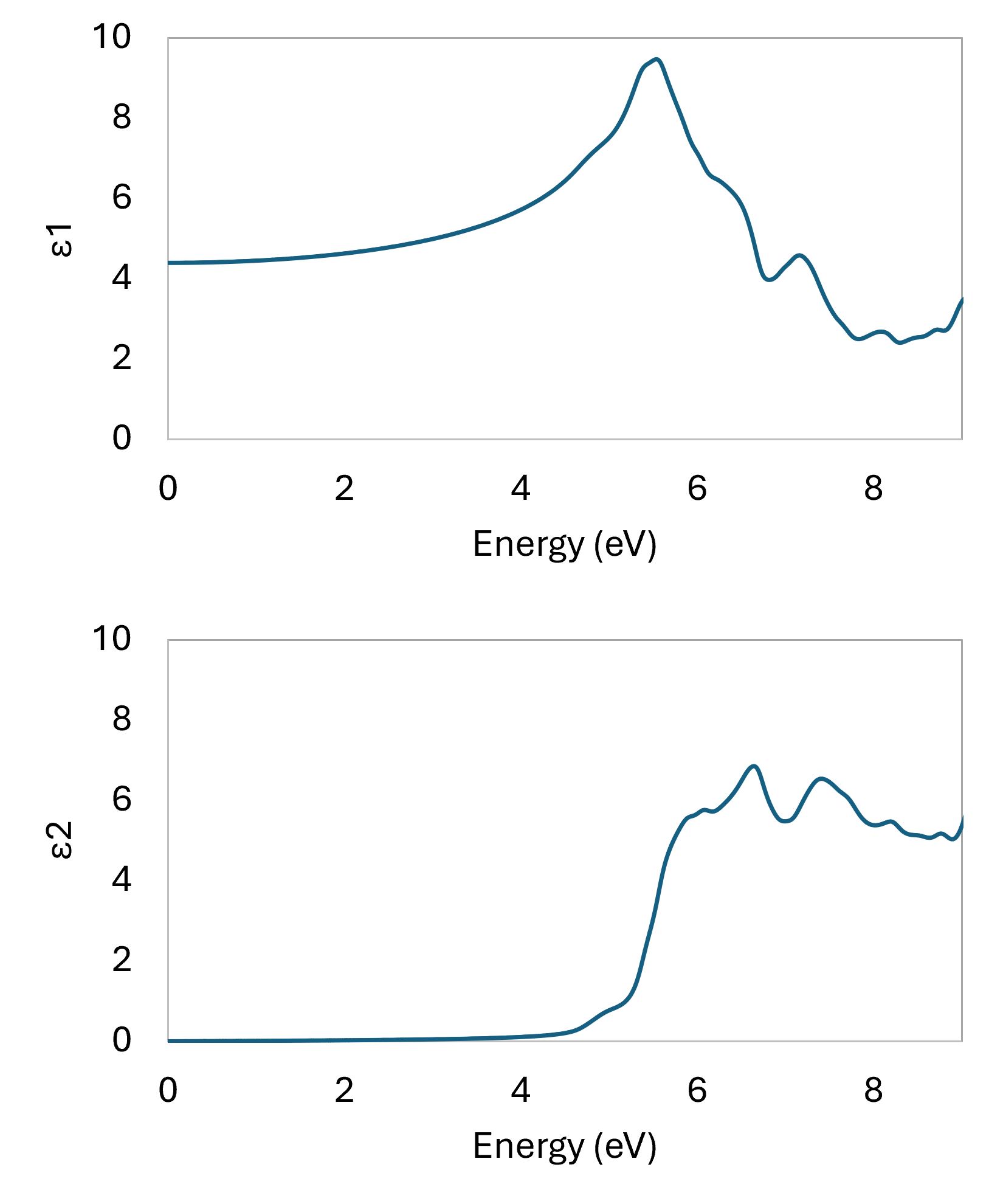}
    \caption{Real ($\epsilon_1$) and imaginary ($\epsilon_2$) parts of the frequency-dependent dielectric function for bulk \ce{ThO2} computed with $G_0W_0$+BSE.}
    \label{fig:BSE dielectric function}
\end{figure}

We can validate our method against published experimental data by computing the dielectric function with $G_0W_0$+BSE.
Our computed data, shown in Fig.~\ref{fig:BSE dielectric function}, are an excellent match to the spectroscopic data in Ref.~\cite{Mock2019}.

\begin{figure}[htb]
    \centering
    \includegraphics[width=0.5\linewidth]{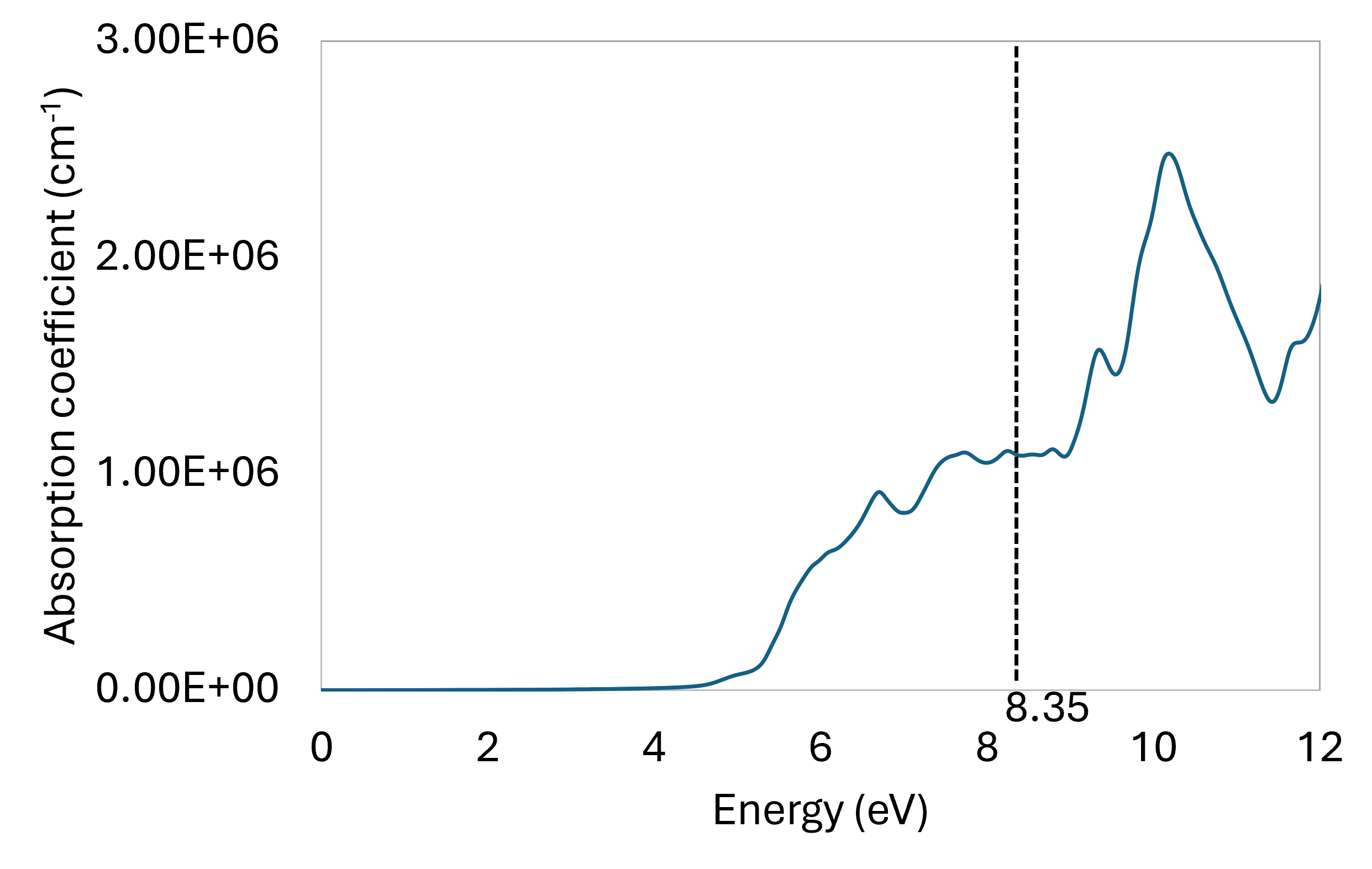}
    \caption{Absorption spectrum of bulk \ce{ThO2} computed with $G_0W_0$+BSE}
    \label{fig:ThO2 G0W0+BSE absorption spectrum}
\end{figure}

We also compute the absorption spectrum of \ce{ThO2} with $G_0W_0$+BSE, as shown in Fig.~\ref{fig:ThO2 G0W0+BSE absorption spectrum}.
The value at the nuclear transition energy is $1\times10^6$ cm$^{-1} \equiv 0.1$ nm$^{-1}$, in agreement with ellipsometric measurements~\cite{tho2_abs_ellipsometry}.

The absorption spectrum plotted in the main text was computed using a 500 eV plane-wave cut-off and the O\_GW pseudopotential, though later testing revealed that these changes did not have a noticeable effect on the calculated absorption spectrum.
Absorption spectra and dielectric functions were processed from VASP output with VASPKIT.

\section{Isomer shift}
\label{SI:Sec:IsomerShift}

The isomer shift originates from differences in the nuclear charge distribution between the ground and excited nuclear states. The value of the shift depends on the local electronic environment of \thor{}. Here we follow the formalism~\cite{perera2025-isomer-shift-Th229} that  combines relativistic many-body atomic structure methods with periodic DFT to evaluate the isomer shifts in \thox{}. The scalar relativistic periodic DFT reproduces valence‑band properties while relativistic atomic‑structure methods captures the essential core‑electron relaxation effects. The valence-band (VB) contribution to the isomer shift in \thor{} solid-state hosts is expressed as~\cite{perera2025-isomer-shift-Th229}
\begin{equation}
\delta E_{\mathrm{iso}}^{\mathrm{VB}}
 \;=\;
 \sum_{\ell}
 \mathrm{IPDOS}_{\ell}\,
 \delta\varepsilon_{\ell}^{\mathrm{iso}}\!\left(\mathrm{Th}^{3+}\right) \, ,
\label{Eq:VB-iso-konec}
\end{equation}
where $\mathrm{IPDOS}_{\ell}$ denotes the integrated projected (on \thor{}) valence band density of states for angular‑momentum  \(\ell\), and
$\delta\varepsilon_{\ell}^{\mathrm{iso}}\!\bigl(\mathrm{Th}^{3+}\bigr)$ are the isomer-shift for the lowest-energy valence orbitals of Th$^{3+}$ of angular‑momentum  \(\ell\). The VB isomer shift is to be added to the isomer shift in \thor$^{4+}$ ion; this contribution remains constant across a wide range of materials.
Table~\ref{tab:isomer-combined} presents the calculated isomer shifts for \thox{} for three different methods. Compared to the PBE and MBJ~\cite{RN489,RN490} methods, the $G_0W_0$ method includes self-energy correction and, as discussed earlier, we consider it of a higher quality.

\begin{table}[h]
\caption{\label{tab:isomer-combined}%
Integrated partial densities of states, $\mathrm{IPDOS}_{\ell}$, and the
products, $\delta E_{\mathrm{iso},\ell}^{\mathrm{VB}}$ = $\mathrm{IPDOS}_{\ell}\,
\delta\varepsilon_{\ell}^{\mathrm{iso}}\!\bigl(\mathrm{Th}^{3+}\bigr)$ obtained with the three electronic-structure methods.  The isomer shift of the solid-state
Th clock, $\delta E_{\mathrm{iso}}^{\mathrm{VB}}$ (bottom
row), is the summation over the orbital, $\ell$
defined in Eq.~(\ref{Eq:VB-iso-konec}). 
}
\begin{ruledtabular}
\begin{tabular}{lcc|cc}
\multirow{2}{*}{$\ell$} &
\multicolumn{2}{c|}{$G_0W_0$} &
\multicolumn{2}{c}{MBJ} \\
 & 
 IPDOS$_\ell$ &
 \( \delta E_{\mathrm{iso},\ell}^{\mathrm{VB}} \) (a.u.) &
 IPDOS$_\ell$ &
 \(\delta E_{\mathrm{iso},\ell}^{\mathrm{VB}}\) (a.u.) \\
\hline
$s$ &  0.119 & \(+1.62\times10^{-8}\) & 0.078 & \(+1.07\times10^{-8}\) \\
$p$ &  0.435 & \(-1.69\times10^{-9}\) & 0.408 & \(-1.58\times10^{-9}\) \\
$d$ &  0.759 & \(-2.42\times10^{-8}\) & 0.688 & \(-2.19\times10^{-8}\) \\
$f$ &  0.534 & \(-4.25\times10^{-8}\) & 0.432 & \(-3.43\times10^{-8}\) \\
\hline
\( \delta E_{\mathrm{iso}}^{\mathrm{VB}} \) (MHz) &

\multicolumn{2}{c|}{\(-343\)} &
\multicolumn{2}{c}{\(-310\)} \\
\end{tabular}
\end{ruledtabular}
\end{table}

Table~\ref{tab:nu_combined} presents the calculated isomer shifts \( \delta E_{\mathrm{iso}}^{\mathrm{VB}} \), the corresponding nuclear-clock frequencies \( \nu \) of \(^{229}\)Th, and their offsets \( \Delta \nu \) relative to the \(\text{ThO}_2\) reference for a range of solid-state hosts.

\begin{table}[h]
\centering
\caption{\label{tab:nu_combined}%
Nuclear‐clock frequencies \( \nu \) obtained from the isomer shifts
\( \delta E_{\mathrm{iso}}^{\mathrm{VB}} \) and their offsets
\( \Delta \nu \) relative to
\(\text{ThO}_2\) computed with MBJ.
The absolute frequencies are referenced to the free‐ion value
\( \nu(\mathrm{Th}^{4+}) = 2\,020\,407\,648(70)\,\mathrm{MHz}\) taken from
Ref.~\cite{perera2025-isomer-shift-Th229}.
}
\begin{ruledtabular}
\begin{tabular}{lccc}
\textbf{Host} &
\( \delta E_{\mathrm{iso}}^{\mathrm{VB}} \) (MHz) &
\( \nu \) (MHz) &
\( \Delta \nu \) (MHz) \\
\hline
ThO\(_2\)     & $-310$ & $2\,020\,407\,338(70)$ & \phantom{$-$}0 \\
CaF\(_2\) F--$90^\circ$--F, Ref.~\cite{perera2025-isomer-shift-Th229}      & $-264$ & $2\,020\,407\,384(70)$ & \phantom{$-$}+46 \\
\lisaf{}, Ref.~\cite{perera2025-isomer-shift-Th229} & $-234$ & $2\,020\,407\,414(70)$ & \phantom{$-$}+76 \\
\end{tabular}
\end{ruledtabular}
\end{table}